\documentclass[11pt]{article}
\usepackage{latexsym}
\usepackage{amsmath} % it subsumes: amstext, amsbsy, amsopn

\begin{document}

\title{{\bf A New Look on the Electromagnetic Duality. \\
            Suggestions and Developments}}
\author{{\bf Stoil Donev}\footnote{e-mail:
 sdonev@inrne.bas.bg} \\ Institute for Nuclear Research and Nuclear
Energy,\\ Bulg.Acad.Sci., 1784 Sofia, blvd.Tzarigradsko chausee 72\\
Bulgaria\\}

\date{}
\maketitle

\begin{abstract}
In this paper a new look on the electro-magnetic duality is presented and
appropriately exploited.  The duality analysis in the nonrelativistic and
relativistic formulations is shown to lead to the idea the mathematical model
field to be a differential form valued in the 2-dimensional vector space
${\cal R}^2$. A full ${\cal R}^2$ covariance is achieved through introducing
explicitly the canonical complex structure ${\cal I}$ of ${\cal R}^2$ in the
nonrelativistic equations. The connection of the relativistic Hodge $*$ with
${\cal I}$ is shown and a complete coordinate free relativistic form of the
equations and the conservative quantities is obtained.  The duality symmetry
is interpreted as invariance of the conservative quantities and conservation
equations.

\end{abstract}

\section{Introduction}
The recent developments in superstring and brane theories show that the
concept of duality plays and probably will play more and more fundamental
role in theoretical and mathematical physics. In today's view duality has a
very general sense and, shortly speaking, it is considered and understood as
availability of (at least) two different ways to describe the same physical
phenomena, thus, the set of all descriptions of the same physical
situation factors over the equivalent ones and, in this way, the essential
differences and peculiarities of the inequivalent descriptions are explicitly
accounted for. This general approach requires a strong insight and a detailed
knowledge of modern quantum and nonquantum field theories.

The roots of duality in field theory, as well as the roots of the whole field
theory, can be found in classical electrodynamics, i.e.  in Maxwell
equations.  Therefore, a detailed and thorough understanding of the duality
nature of classical electrodynamics seems to be a very important first step
in getting well acquainted with the duality notion.  At first sight the
electromagnetic duality seems quite simple and not much promising theoretical
tool, but a closer look reveals the opposite. In fact, as we shall see later,
a careful study of the duality brings us to the conclusion that the adequate
mathematical model object of the electromagnetic field should have a more
complicated structure than just the couple $(\boldsymbol{E},\boldsymbol{B})$,
or ${\bf F}_{\mu\nu}$. As it is well known [1], one of the most frequently
used ways of introducing duality makes use of complex electro-magnetic
vectors $(\boldsymbol{E}+i\boldsymbol{B})$ and transformation of the kind
$(\boldsymbol{E}+i\boldsymbol{B})\rightarrow e^{i\varphi}({\bf
E}+i\boldsymbol{B})$, but such a formal step seems closer to a peculiarity
notice than to reveal an important feature of the theory, while the real
presence of the ${\cal R}^2$-complex structure inside the electromagnetic
theory seems to stay not fully understood as its structure property.
Following such a line of consideration many questions stay unanswered, e.g.
why {\it complex} vector fields on {\it real} manifolds; why
$i\boldsymbol{B}$ and not $i\boldsymbol{E}$; why in the relativistic
formulation $i=\sqrt{-1}$ disappears and, instead, a complex structure
(through the Hodge $*$) in the 6-dimensional module of 2-forms on the
Minkowski space appears, etc.

As we shall see, an adequate (nonrelativistic or relativistic) formulation of
classical electrodynamics really needs complex structure, but {\it not}
complex vectors on real manifolds.  The duality transformations turn to be
those linear isomorphisms of the standard complex structure ${\cal I}$ in
${\cal R}^2$, which are also isometries of the standard metric in ${\cal
R}^2$.  In our approach duality transformations appear as closer related to
the invariance of the conserved quantities of the theory than to the
invariance properties of the equations.  And this should be expected since we
can treat them as natural extension of the usual space-time isometries,
which, as we know, are basic tools in formulating and computing the
conservative characteristics of any physical field.

The first man who noticed the duality properties of electromagnetic equations
was Heaviside [2]. A further development of Heaviside's notice was given
later by Larmor [3]. A more detailed study of electromagnetic duality was
made by Rainich [4] in the frame of General Relativity. A comparatively
complete presentation may be found in the extensive paper of Misner and
Wheeler [5] also in the frame of General Relativity in connection with their
attempt to geometrize classical physics and to give topological
interpretation of charges. Electromagnetic duality has always been in sight
of those trying to introduce magnetic charges and currents in the theory,
[6],[7]. In [8] one may find duality considerations in the frame of nonlinear
electrodynamics of continuous media.  A formal generalization for p-forms is
given in [9].  A modern consideration of electromagnetic duality, directed
to superstring and brane theories may be found in [1]; the possible
nonabelian generalizations are considered in [10].

In this paper we pursue three main purposes. First, we are going to give a
brief nonrelativistic and relativistic reviews of what is usually called
electromagnetic duality in vacuum and in presence of electric and magnetic
charges, without referring to other physical theories.  Second, we shall
present a new understanding of this duality, leading to a new and, in our
view, more adequate mathematical nature of the (nonrelativistic and
relativistic) model objects plus complex structure. Finally, we shall
represent classical electrodynamics entirely in terms of the new
nonrelativistic and relativistic mathematical model objects.

\vskip 0.3cm

\section{Electromagnetic Duality}
\vskip 0.3cm
{\bf 2.1 Nonrelativistic consideration.} We consider first the pure field
Maxwell equations
\begin{equation}
{\rm rot}{\boldsymbol{E}}+
\frac 1c      \frac{\partial {\boldsymbol{B}}} {\partial t}=0
, \quad {\rm div}\boldsymbol{B}=0,             %1%
\end{equation}
\begin{equation}
{\rm rot}\boldsymbol{B} -
	\frac 1c \frac{\partial {\boldsymbol{E}}} {\partial t}=0 ,
\quad {\rm div}{\boldsymbol{E}}=0.                                 %2%
\end{equation}

First we note, that because of the linearity of these equations if
$({\boldsymbol{E}}_i,\boldsymbol{B}_i), i=1,2,...$  are a collection of solutions,
then every couple of linear combinations of the form
\begin{equation}
{\boldsymbol{E}}=a_i{\boldsymbol{E}}_i,\ \boldsymbol{B}=b_i\boldsymbol{B}_i   %3%
\end{equation}
(sum over the repeated $i=1,2,...$) with arbitrary constants $(a_i,b_i)$
gives a new solution.

We also note, that in the static case the pure field Maxwell equations
reduce to:
\begin{equation}
*{\bf d}{\boldsymbol{E}}=0, *{\bf d}\boldsymbol{B}=0,
*{\bf d}*\boldsymbol{B}=0, *{\bf d}*{\boldsymbol{E}}=0,                          %4%
\end{equation}
where $*$ is the euclidean Hodge $*$-operator, $\boldsymbol{d}$ is the
exterior
derivative and the vector fields ${\boldsymbol{E}}$ and $\boldsymbol{B}$ are
identified with
the corresponding 1-forms through the euclidean metric $g$.  The substitution
${\boldsymbol{E}}\rightarrow *{\boldsymbol{E}},\ \boldsymbol{B}\rightarrow *\boldsymbol{B}$, because of the
relation $*^2=\pm id$, turns the first equation of (4) into the fourth one
and vice versa, the fourth - into the first one; also, the second equation is
turned into the third one, and vice versa, the third - into the second one.
Hence, in this special case we can talk about $*$-symmetry of the equations.

The important observation made by Heaviside [1], and later considered
by Larmor [2], is that the substitution
\begin{equation}
{\boldsymbol{E}}\rightarrow -\boldsymbol{B},\quad \boldsymbol{B}\rightarrow {\boldsymbol{E}}      %5%
\end{equation}
transforms the first couple (1) of the pure field Maxwell equations into the
second couple (2), and, vice versa, the second couple (2) is transformed into
the first one (1). This symmetry transformation (5) of the pure field Maxwell
equations is called {\it special duality transformation}, or
SD-transformation.  It clearly shows that the electric and magnetic
components of the pure electromagnetic field are interchangeable and the
interchange (5) transforms solution into solution. In the transformed
solution the magnetic component is the former electric component and vice
versa, i.e. the electric component may be considered as magnetic if needed,
and then the magnetic component should be considered as electric. This
feature of the pure electromagnetic field reveals its {\it dual} nature.

It is important to note that the SD-transformation (5) does not
change the energy density $8\pi{\bf w}={\boldsymbol{E}}^2+\boldsymbol{B}^2$,
the Poynting vector $4\pi\boldsymbol{S}=c({\boldsymbol{E}}\times\boldsymbol{B})$ , and the
(nonlinear) Poynting relation
\[
\frac {\partial}{\partial t} \frac {{\boldsymbol{E}}^2+\boldsymbol{B}^2}{8\pi}=
-{\rm div}\boldsymbol{S}.
\]
Hence, from energy-momentum point of view two dual, in the sense of (5),
solutions are indistinguishable.

Note that the substitution (5) may be considered as a transformation of the
following kind:
\begin{equation}
(\boldsymbol{E},\boldsymbol{B})
	%\left\| \matrix{0   &1\cr -1  &0\cr} \right\|=      %6%
	\begin{Vmatrix}
0   &1
\\
-1  &0
	\end{Vmatrix}
=
(-\boldsymbol{B},\boldsymbol{E}).
\end{equation}
The following question now arises naturally: do there exist constants
$(a,b,m,n)$, such that the linear combinations
\begin{equation}
{\boldsymbol{E}}'=a{\boldsymbol{E}}+m\boldsymbol{B},\ \boldsymbol{B}'=b{\boldsymbol{E}}+n\boldsymbol{B},       %7%
\end{equation}
or in a matrix form
\begin{equation}
({\boldsymbol{E}}',\boldsymbol{B}')=({\boldsymbol{E}},\boldsymbol{B})
%\left\|\matrix{a   &b\cr m  &n\cr} \right\|
	\begin{Vmatrix}
a  & b
\\
m  & n
	\end{Vmatrix}
=
(a{\boldsymbol{E}}+m\boldsymbol{B},b{\boldsymbol{E}}+n\boldsymbol{B}),                         %8%
\end{equation}
form again a vacuum solution?
Substituting $\boldsymbol{E}'$ and $\boldsymbol{B'}$ into Maxwell's vacuum
equations we see that the answer to this question is affirmative iff $m=-b,
n=a$, i.e. iff the corresponding matrix $S$ is of the form \begin{equation}
S=
%\left\|\matrix{a   &b\cr -b  &a\cr} \right\|
	\begin{Vmatrix}
a	& b
\\
-b	& a                                   %9%
	\end{Vmatrix}
.
\end{equation}
The new solution
will have now energy density ${\bf w}'$ and momentum density $\boldsymbol{S}'$ as
follows:
\[
{\bf w}'=\frac {1}{8\pi}\biggl({\boldsymbol{E}'}^2 +
			{\boldsymbol{B}'}^2\biggr)=
\frac {1}{8\pi}(a^2 + b^2)\biggl({\boldsymbol{E}}^2 + \boldsymbol{B}^2\biggr),
\]
\[\boldsymbol{S}'=(a^2+b^2)\frac {c}{4\pi} {\boldsymbol{E}}\times
			\boldsymbol{B}.
\]
Obviously, the new and the old solutions will have the same energy and
momentum if $a^2+b^2 =1$, i.e. if the matrix $S$ is {\it unimodular}.
In this case we may put
$a=\cos \alpha$ and $b=\sin \alpha$, where $\alpha=const$,
so transformation (8) becomes
\begin{equation}
\tilde{\boldsymbol{E}}={\boldsymbol{E}}\cos \alpha -
				\boldsymbol{B}\sin \alpha,\  %10%
\tilde{\boldsymbol{B}}={\boldsymbol{E}}\sin \alpha + \boldsymbol{B}\cos
\alpha.
	\end{equation}

Transformation (10) is known as {\it electromagnetic duality transformation},
or D-transformation. It has been a subject of many detailed studies in
various aspects and contexts [1],[3]-[9].  It also has greatly
influenced some modern developments in non-Abelian Gauge theories, as well as
some recent general views on duality in field theory, esp. in superstring and
brane theories (classical and quantum). In the next section we shall study
this transformation from a new point of view, following the idea that
$(\boldsymbol{E},\boldsymbol{B})$ are two vector components of one
mathematical object having some more complicated nature.

From physical point of view a basic feature of the D-transformation (10) is,
that the difference between the electric and magnetic fields becomes
non-essential: we may superpose the electric and the magnetic vectors, i.e.
vector-components, of a general electromagnetic field to obtain new
solutions.  From mathematical point of view we see that Maxwell's equations
in vacuum, besides the usual linearity (3) mentioned above, admit also
"cross"-linearity, i.e.  linear combinations of ${\boldsymbol{E}}$ and $\boldsymbol{B}$ of a
definite kind define new solutions: with every solution
$({\boldsymbol{E}},\boldsymbol{B})$
of Maxwell's vacuum equations a 2-parameter family of solutions can be
associated by means of linear transformations given by matrices of the kind
(9). If these matrices are unimodular, i.e. when $a^2+b^2=1$, then all
solutions of the family have the same energy and momentum. In other words,
the space of all solutions to the pure field Maxwell equations factors over
the action (8) of the group of linear maps $S:{\cal R}^2\rightarrow{\cal
R}^2$ represented by matrices of the kind (9).

It is well known that matrices $S$ of the kind (9) do not change the
canonical complex structure ${\cal I}$ in ${\cal R}^2$: we recall that if the
canonical basis of ${\cal R}^2$ is denoted by $(\varepsilon^1,\varepsilon^2)$
then ${\cal I}$ is defined by ${\cal I}(\varepsilon^1)=\varepsilon^2$, ${\cal
I}(\varepsilon^2)=-\varepsilon^1$, so if $S$ is given by (9) we have:
$S.{\cal I}.S^{-1}={\cal I}$.  Hence, {\bf the electromagnetic
D-transformations} (10) {\bf coincide with the unimodular symmetries of the
canonical complex structure} ${\cal I}$ {\bf in} ${\cal R}^2$.  This
important in our view remark clearly points out that the canonical complex
structure ${\cal I}$ in ${\cal R}^2$ should be an {\bf essential element} of
classical electromagnetic theory, so we should not forget about it and in no
way neglect it. Moreover, in my opinion, we {\bf must find an appropriate way
to introduce} ${\cal I}$ {\bf explicitly} in the theory, and if this is done
we should expect a covariance with respect to general nondegenerate
transformations (7), not just covariance with respect to transformations
(9), which are usually considered as symmetries .

In presence of electric current $\boldsymbol{j}_e$ and electric charges $\rho_e$
Maxwell equations (1)-(2) lose this D-symmetry.  In order
to retain it {\it magnetic charges} with density $\rho_m$ and magnetic
currents $\boldsymbol{j}_m=\rho_m\boldsymbol{v}$ are usually introduced, and of course, the
Lorentz force is correspondingly modified. The new system of equations looks
as follows:
\begin{equation}
{\rm rot}{\boldsymbol{E}}+\frac 1c \frac{\partial \boldsymbol{B}}{\partial t}=-        %11%
 \frac {4\pi}{c}\boldsymbol{j}_m, \quad {\rm div}\boldsymbol{B}=4\pi\rho_m,
\end{equation}
\begin{equation}
{\rm rot}\boldsymbol{B} -
		\frac 1c \frac{\partial {\boldsymbol{E}}}{\partial t}=
\frac{4\pi}{c}\boldsymbol{j}_e,
\quad  {\rm div}{\boldsymbol{E}}=4\pi\rho_e,                                   %12%
\end{equation}
\begin{equation}
\mu\nabla_{\boldsymbol{v}} \boldsymbol{v}=\rho_e \boldsymbol{E}+
\frac 1c (\boldsymbol{j}_e\times \boldsymbol{B}) + \rho_m \boldsymbol{B}
-\frac 1c (\boldsymbol{j}_m\times {\boldsymbol{E}}),    %13%
\end{equation}
where $\mu$ is the mass density of the particles and it is assumed that
particles do not vanish and do not arise.
Now the whole, or extended, D-transformation looks in the following way:
\begin{equation}
\tilde{\boldsymbol{E}}={\boldsymbol{E}}\cos \alpha -
				\boldsymbol{B}\sin \alpha,\  %14%
\tilde{\boldsymbol{j}}_e=\tilde{q}_e\boldsymbol{v}=
\boldsymbol{j}_e\cos \alpha-\boldsymbol{j}_m\sin \alpha
=(\rho_e\cos \alpha - \rho_m\sin \alpha)\boldsymbol{v}
\end{equation}
\begin{equation}
\tilde{\boldsymbol{B}}={\boldsymbol{E}}\sin \alpha +
			 \boldsymbol{B}\cos \alpha,\
\tilde{\boldsymbol{j}}_m=\tilde q_m\boldsymbol{v}=
\boldsymbol{j}_e\sin \alpha+\boldsymbol{j}_m\cos \alpha=               %15%
(\rho_e\sin \alpha + \rho_m\cos \alpha)\boldsymbol{v}.
\end{equation}
Hence, the 1-parameter family of transformations (14-15)
is a symmetry of the system (11-13).

The corresponding considerations concerning one particle carrying electric
and magnetic charges may be found in [6]. We shall omit this simple case.

Finally we note that D-transformations change the two well known
invariants: $I_1=(\boldsymbol{B}^2-{\boldsymbol{E}}^2)$ and $I_2=2{\boldsymbol{E}}.\boldsymbol{B}$ in the
following way:
\begin{equation}
\tilde{I_1}=\tilde{\boldsymbol{B}}^2-\tilde{\boldsymbol{E}}^2=
(\boldsymbol{B}^2-{\boldsymbol{E}}^2)\cos 2\alpha+
			2{\boldsymbol{E}}.\boldsymbol{B}\sin 2\alpha=
I_1\cos 2\alpha+I_2\sin 2\alpha,                %16%
\end{equation}
\begin{equation}
\tilde{I_2}= 2\tilde{\boldsymbol{E}}.\tilde{\boldsymbol{B}}=
({\boldsymbol{E}}^2-\boldsymbol{B}^2)\sin 2\alpha+
			2{\boldsymbol{E}}.\boldsymbol{B}\cos 2\alpha=
-I_1\sin 2\alpha+I_2\cos 2\alpha.                  %17%
\end{equation}
It follows immediately that
\[
\tilde{I_1}^2+\tilde{I_2}^2=I_1^2+I_2^2,
\]
i.e. the sum of the squared invariants is a D-invariant.

\vskip 0.3cm
\noindent
{\bf 2.2 Relativistic consideration.} Recall Maxwell's pure field equations in
relativistic form
\begin{equation}
{\bf d}{\bf F}=0,\ {\bf d}*{\bf F}=0.                %18%
\end{equation}
The Hodge $*$-operator is defined by the relation
$$
\alpha\wedge *\beta=-\eta(\alpha,\beta)\sqrt{|det(\eta_{\mu\nu})|}dx\wedge
dy\wedge dz\wedge d\xi,
$$
where $\alpha$ and $\beta$ are a $p$-forms on the Minkowski spacetime $M$,
$\xi=ct$ is the time coordinate, and the Minkowski metric
$\eta$ has signiture $(-,-,-,+)$.

We note first 2 simple symmetries of (18).

$1^o$. The transformation ${\bf F}\rightarrow *{\bf F}$ keeps the system (18)
the same.  This follows from the property $*(*{\bf F})=-{\bf F}$ of the
$*$-operator when restricted on 2-forms.

$2^o$. Any {\it conformal} change of the Minkowski metric $\eta\rightarrow
f^2 \eta$, where $f$ is everywhere different from zero function on the
Minkowski space $M$, keeps the restriction of the Hodge $*$ to 2-forms on $M$
the same, so (18) is conformally invariant.

Recalling the explicit form of ${\bf F}$ and $*{\bf F}$ in canonical
coordinates
\begin{eqnarray}
{\bf F}=\boldsymbol{B}_3dx\wedge dy
-\boldsymbol{B}_2dx\wedge dz +\boldsymbol{B}_1dy\wedge dz +
{\boldsymbol{E}}_1dx\wedge d\xi +{\boldsymbol{E}}_2dy\wedge d\xi +
{\boldsymbol{E}}_3dz\wedge d\xi                                     %19%
\end{eqnarray}
\begin{eqnarray}
*{\bf F}={\boldsymbol{E}}_3dx\wedge dy -{\boldsymbol{E}}_2dx\wedge dz
+{\boldsymbol{E}}_1dy\wedge dz -
\boldsymbol{B}_1dy\wedge dz -\boldsymbol{B}_2dy\wedge d\xi
-\boldsymbol{B}_3dz\wedge d\xi    %20%
\end{eqnarray}

we see that $*$ replaces $\boldsymbol{E}$ with $-\boldsymbol{B}$ and
$\boldsymbol{B}$ with ${\boldsymbol{E}}$, i.e.  the action of $*$ gives the
SD-transformation. On the other hand an extended SD-transformation may be
introduced by
\[
({\bf F},*{\bf F})
%\left\|\matrix{0 &-1\cr 1  &0\cr} \right\|
\begin{Vmatrix} 0	& -1 \\ 1	& 0	\end{Vmatrix}
=(*{\bf F},-{\bf F}).
\]
We note the difference: the $*$-operator transforms a $p$-form $\beta$ into a
$(4-p)$-form $*\beta$, while the above SD-transformation transforms a
{\it couple} of forms to another {\it couple} of forms.

As in the nonrelativistic case, this SD-transformation is readily extended to
the full  D-transformation
\[
({\bf F},*{\bf F})\rightarrow({\cal F},\tilde{\cal F})=({\bf F},*{\bf F})
%\left\|\matrix{\cos \alpha   &-\sin \alpha \cr \sin \alpha
%	&\cos \alpha\cr} \right\|
	\begin{Vmatrix}
\cos\alpha	& -\sin\alpha
\\
\sin\alpha	& \cos\alpha
	\end{Vmatrix}
,
\]
i.e.
\begin{equation}
{\cal F}={\bf F}\cos \alpha+*{\bf F}\sin \alpha,\          %21%
\tilde{\cal F}=-{\bf F}\sin \alpha+*{\bf F}\cos \alpha.
\end{equation}
The above transformations transform solutions to solutions. Moreover,
{\it in contrast} to the nonrelativistic case, where linear combinations of
${\boldsymbol{E}}$ and $\boldsymbol{B}$ of {\it special} kind transform solutions to
solutions, here every linear combination of ${\bf F}$ and $*{\bf F}$, i.e. a
transformation of the kind
\begin{equation}
{\cal F}_g=a{\bf F}+b*{\bf F},\
\tilde{\cal F}_g=m{\bf F}+n*{\bf F},           %22%
\end{equation}
where the subscribe $g$ means "general", with arbitrary constants $(a,b,m,n)$
gives again a solution. As we shall see this is because some of the special
properties of $S$ are now hidden in the $*$-operator through the pseudometric
$\eta$, and the components ${\bf F}$ and $*{\bf F}$ are interrelated.

It is important to note that transformation (21) keeps the energy-momentum
tensor
\begin{equation}
{\bf Q}_\mu^\nu=\frac {1}{4\pi}\biggl[\frac{1}{4}
{\bf F}_{\alpha\beta}{\bf F}^{\alpha\beta}\delta_\mu^\nu-
{\bf F}_{\mu\sigma}{\bf F}^{\nu\sigma}\biggr]=
\frac {1}{8\pi}\biggl[-{\bf F}_{\mu\sigma}{\bf F}^{\nu\sigma}-
(*{\bf F})_{\mu\sigma}(*{\bf F})^{\nu\sigma}\biggr],                  %23%
\end{equation}
\noindent
and its divergence
\begin{equation}
\nabla_\nu {\bf Q}_\mu^\nu=-
\frac {1}{4\pi}\biggl[{\bf F}_{\mu\nu}(\nabla_\sigma {\bf F}^{\sigma\nu})+
(*{\bf F})_{\mu\nu}(\nabla_\sigma (*{\bf F})^{\sigma\nu})\biggr] %24%
\end{equation}
the same: ${\bf Q}_\mu^\nu({\bf F})=Q_\mu^\nu({\cal F})$,
$\nabla_\nu {\bf Q}_\mu^\nu({\bf F})=
\nabla_\nu {\bf Q}_\mu^\nu({\cal F})$.

% while the general linear transformation (22) changes it.

We see that the relativistic formulation of Maxwell theory naturally admits
general ${\cal R}^2$-covariance as far as transformations (22) are implied
to act on two 2-forms of the kind $({\bf F},*{\bf F})$. As for the
D-transformations, they are closely related to the symmetries of the
energy-momentum quantities and relations.

The two quantities
\[
(4\pi)^2Q_{\mu\nu}Q^{\mu\nu}=I_1^2+I_2^2,\
(4\pi)^2Q_{\mu\sigma}Q^{\nu\sigma}=\frac14(I_1^2+I_2^2)\delta_\mu^\nu
\]
also enjoy the D-invariance. We note also that the eigen values of
${\bf Q}_\mu^\nu$ are D-invariant, as it should be, while the eigen
values of ${\bf F}$ and $*{\bf F}$, given respectively by
\[
\lambda_{1,2}=\pm \sqrt{-\frac12 I_1 +\frac12 \sqrt{I_1^2+I_2^2}} ,\quad
\lambda_{3,4}=\pm \sqrt{-\frac12 I_1 -\frac12 \sqrt{I_1^2+I_2^2}} ,
\]
\[
\lambda^*_{1,2}=\pm \sqrt{\frac12 I_1 +\frac12 \sqrt{I_1^2+I_2^2}} ,\quad
\lambda^*_{3,4}=\pm \sqrt{\frac12 I_1 -\frac12 \sqrt{I_1^2+I_2^2}}
\]
are not  D-invariant. Only when $I_1=I_2=0$, the so called {\it null
field case}, the eigen values of ${\bf F}$  and $*{\bf F}$ are D-invariant
since in this case they are zero.

The relativistic generalization of equations (11)-(13) is
\begin{equation}
\nabla_\nu {\bf F}^{\nu\mu}=-4\pi j^\mu_e,             %25%
\end{equation}
\begin{equation}
\nabla_\nu (*{\bf F})^{\nu\mu}=-4\pi J^\mu_m,            %26%
\end{equation}
\begin{equation}
\mu c^2 u^\nu\nabla_\nu u_\mu=
-{\bf F}_{\mu\nu} j^\nu_e -(*{\bf F})_{\mu\nu} J^\nu_m,        %27%
\end{equation}
where $\mu$ is the invariant mass density, $j^\mu_e=\rho_e u^\mu$,
$J^\mu_m=(\boldsymbol{J}_m, J^4)=(-\boldsymbol{j}_m, -\rho_m)$, and $u^\mu$ is the
4-velocity vector field. The generalized Lorentz force (13) is obtained from
(27) after raising the index $\mu$ through $\eta^{\mu\nu}$.

The above system (25)-(27) enjoys the following symmetry transformation:
\[
{\bf F}\rightarrow *{\bf F};\ j_e\rightarrow J_m; \ J_m\rightarrow -j_e.
\]
This invariance is a particular case of the more general invariance
transformation given by relations (21) plus
\[
j'_e=j_e\cos \alpha - J_m\sin \alpha,\
J'_m=j_e\sin \alpha + J_m\cos \alpha,
\]
which is readily checked. It should be noted that these invariances make use
of the property $*^2=-id$ of the restricted to 2-forms Hodge $*$-operator, and
of the {\it constancy} of the phase angle $\alpha$ in the D-transformation.

Finally we'd like to note the different physical sense of equation (27)
compare to equations (25)-(26): equation (27) equalizes changes
of energy-momentum densities, it is a direct differential form of the
energy-momentum balance relation between the field and the particles,
carrying mass and electric and magnetic charge; equations (25)-(26) are
relativistic forms of the differential equivalents of the time changes of the
flows through 2-surfaces of ${\boldsymbol{E}}$ and $\boldsymbol{B}$,
moreover, these two equations identify quantities of {\it different physical
nature}:  it is hard to believe that differentiating field functions (the
left hand side) we could obtain currents of charged mass particles (the right
hand side).

\section{The Suggestion and Developments}

{\bf 3.1 \ Nonrelativistic formulation}. We summurize some of the D-features
of the field description through Maxwell equations.

1. The D-invariance of the field equations is a mathematical representation
of the dual {\it electro-magnetic} $({\boldsymbol{E}},\boldsymbol{B})$-nature of the field.
This dual nature is explicitly seen in the nonrelativistic form of Maxwell's
equations: ${\boldsymbol{E}}$ and $\boldsymbol{B}$ depend on each other but they can be
always distinguished from each other.

2. All energy-momentum quantities and relations are D-invariant.

3. The D-transformation is represented by a rotation in
a 2-dimensional vector space and acts through superposing the electric
and magnetic components of the field and the corresponding currents.

4. The rotation angle $\alpha$ does not depend on the space-time coordinates.

5. The electro-magnetic duality transformation keeps and emphasises the
field's united nature.

\vskip 0.5cm

The suggestion coming from these notices is that the electromagnetic field,
considered as  {\it one physical object}, has {\it two physically
distinguishable interrelated vector components}, $({\boldsymbol{E}},\boldsymbol{B})$, so the
adequate mathematical model-object must have two vector components and must
admit 2-dimensional linear transformations of its components, in particular,
the 2-dimensional rotations should be closely related to the invariance
properties of the energy-momentum characteristics of the field. But
every 2-dimensional linear transformation requires a "room where to act",
i.e. a 2-dimensional real vector space has to be {\it explicitly pointed
out}. This 2-dimensional space has always been implicitly present inside
the electromagnetic field theory, but has never been
introduced explicitly. Let's introduce it.

\vskip 0.5cm
\noindent
{\it The electromagnetic field is mathematically represented
(nonrelativisticaly) by an ${\cal R}^2$-valued differential 1-form $\omega$,
such that in the canonical basis $(\varepsilon^1,\varepsilon^2)$
in ${\cal R}^2$ the 1-form $\omega$ looks as follows}
\begin{equation}
\omega={\boldsymbol{E}}\otimes \varepsilon^1 +
\boldsymbol{B}\otimes \varepsilon^2. %28%
\end{equation}
\noindent
\vskip 0.5cm
\noindent
{\bf Remark}. In (28), as well as later on, we identify the vector fields and
1-forms on ${\cal R}^3$ through the euclidean metric
and we write, e.g.
$*({\boldsymbol{E}}\wedge\boldsymbol{B})={\boldsymbol{E}}\times\boldsymbol{B}$.
Also, we identify
$({\cal R}^2)^*$ with ${\cal R}^2$ through the euclidean metric.
\vskip 0.5cm
Now we have to present equations (11)-(13) correspondingly, i.e. in terms of
${\cal R}^2$-valued objects.

We begin with the electric $\rho_e$ and magnetic $\rho_m$ densities,
considering them as components of an ${\cal R}^2$ valued function
${\cal Q}$, i.e.
\begin{equation}
{\cal Q}=\rho_e\otimes \varepsilon^1+\rho_m\otimes \varepsilon^2.  %29%
\end{equation}
The two currents $\boldsymbol{j}_e$ and $\boldsymbol{j}_m$, considered as 1-forms, become
components of an ${\cal R}^2$-valued 1-form ${\cal J}$ as follows
\begin{equation}
{\cal J}=\boldsymbol{j}_e\otimes \varepsilon^1+\boldsymbol{j}_m\otimes \varepsilon^2.
\end{equation}                                                  %30%

As we mentioned earlier (p.4), the above assumption (28) requires a general
covariance with respect to transformations in ${\cal R}^2$, so, the
complex structure ${\cal I}$ has to be introduced explicitly in the
equations.  In order to do this we recall that the linear map ${\cal I}:{\cal
R}^2\rightarrow{\cal R}^2$ induces a map ${\cal I}_*:\omega\rightarrow {\cal
I}_*(\omega)={\boldsymbol{E}}\otimes {\cal I}(\varepsilon^1)+ \boldsymbol{B}\otimes {\cal
I}(\varepsilon^2)= -\boldsymbol{B}\otimes \varepsilon^1+ {\boldsymbol{E}}\otimes
\varepsilon^2$.  We recall also that every operator  ${\cal D}$ in the set of
differential forms is naturally extended to vector-valued differential forms
according to the rule ${\cal D}\rightarrow {\cal D}\times id$, and $id$ is
usually omitted. Having in mind the identification of vector fields and
1-forms through the euclidean metric we introduce now ${\cal I}$ in Maxwell's
equations (11)-(13) through $\omega$ in the following way:
\begin{equation}
*{\bf d}\omega-\frac {1}{c} \frac {\partial }{\partial t}{\cal I}_*
(\omega)=                                                              %31%
\frac {4\pi}{c}{\cal I}_*({\cal J}),\ \ \delta \omega =-4\pi {\cal Q},
\end{equation}
where $\delta=*{\bf d}*$ is the codifferential.
Two other equivalent forms of (31) are given as follows:
\[
{\bf d}\omega-*\frac {1}{c} \frac {\partial }{\partial t}{\cal I}_*
(\omega)=
\frac {4\pi}{c}*{\cal I}_*({\cal J}),\ \ \delta \omega =-4\pi {\cal Q},
\]
\[
*{\bf d}{\cal I}_*(\omega)+
\frac {1}{c} \frac {\partial }{\partial t}\omega=-
\frac {4\pi}{c}{\cal J},\ \ \delta \omega =-4\pi {\cal Q}.
\]
\noindent
In order to verify the equivalence of (31) to Maxwell equations (11)-(13)
we compute the marked operations.  For the left-hand side of the first (31)
equation we obtain
\[
*{\bf d}\omega-
\frac {1}{c} \frac {\partial }{\partial t}{\cal I}_*(\omega)=
\left({\rm rot}{\boldsymbol{E}}+\frac1c\frac{\partial \boldsymbol{B}}{\partial t}\right)
\otimes\varepsilon^1+
\left({\rm rot}\boldsymbol{B}-
\frac1c\frac{\partial {\boldsymbol{E}}}{\partial t}\right)\otimes\varepsilon^2,
\]
\noindent
and the right-hand side is
\[
\frac{4\pi}{c}{\cal I}_*({\cal J})=-
\frac{4\pi}{c}\boldsymbol{j}_m\otimes\varepsilon^1+
\frac{4\pi}{c}\boldsymbol{j}_e\otimes\varepsilon^2
\]
\noindent
The second equation $\delta \omega=-4\pi{\cal Q}$ is, obviously,
equivalent to
$$
{\rm div}{\boldsymbol{E}}\otimes \varepsilon^1 +
{\rm div}\boldsymbol{B}\otimes \varepsilon^2=4\pi \rho_e\otimes \varepsilon^1+
4\pi\rho_m\otimes \varepsilon^2
$$
since $\delta=-{\rm div}$. Hence, (31) coincides with (11)-(13).

We shall emphasize once again that according to our general assumption (28)
the field $\omega$ will have different representations in the different bases
of ${\cal R}^2$.
Changing the basis $(\varepsilon^1,\varepsilon^2)$ to
any other basis
$\varepsilon^{1'}=\varphi(\varepsilon^1),
\varepsilon^{2'}=\varphi(\varepsilon^2)$,
means, of course, that in equations (31) the field $\omega$ changes to
$\varphi_*\omega$ and the complex structure ${\cal I}$
changes to $\varphi{\cal I}\varphi^{-1}$. In some sense this means that we
have two fields now: $\omega$ and ${\cal I}$, but ${\cal I}$ is given
beforehand and it is not determined by equations (31). So, in the new basis
the $\mathcal{I}$-dependent equations of (31) will look like
\[
*{\bf d}\varphi_*\omega-
\frac1c\frac{\partial }{\partial t}
(\varphi{\cal I}\varphi^{-1})_*(\varphi_*\omega)=
\frac{4\pi}{c}(\varphi{\cal I}\varphi^{-1})_*(\varphi_*{\cal J}).
\]
If $\varphi$ is a symmetry of ${\cal I}:
\varphi{\cal I}\varphi^{-1}={\cal I}$, then we transform just $\omega$ to
$\varphi_*\omega$.

In order to write down the Poynting energy-momentum balance relation we
recall the product of vector-valued differential forms. Let
$\Phi= \Phi^a\otimes e_a$ and $\Psi= \Psi^b\otimes k_b$ are two differential
forms on some manifold with values in the vector spaces $V_1$ and $V_2$
with bases $\{e_a\}, a=1,...,n$ and $\{k_b\}, b=1,...,m$, respectively.
Let $f:V_1\times V_2\rightarrow W$ is a bilinear map valued in a third vector
space $W$.  Then a new differential form, denoted by $f(\Phi,\Psi)$, on the
same manifold and valued in $W$ is defined by
\[
f(\Phi,\Psi)=\Phi^a\wedge \Psi^b\otimes f(e_a,k_b).
\]
Clearly, if the original forms are $p$ and $q$ respectively,
then the product is a $(p+q)$-form.

Assume now that $V_1=V_2={\cal R}^2$ and the bilinear map is the
exterior product:\linebreak
$\wedge:{\cal R}^2\times {\cal R}^2\rightarrow \Lambda^2({\cal R}^2)$.
Let's compute the expression $\wedge(\omega,{\bf d}\omega)$.
\[
\wedge(\omega,{\bf d}\omega)=\wedge({\boldsymbol{E}}\otimes \varepsilon^1+
\boldsymbol{B}\otimes \varepsilon^2,{\bf d}{\boldsymbol{E}}\otimes \varepsilon^1+
{\bf d}\boldsymbol{B}\otimes \varepsilon^2)=
({\boldsymbol{E}}\wedge{\bf d}\boldsymbol{B}-\boldsymbol{B}\wedge{\bf d}{\boldsymbol{E}})\otimes
\varepsilon^1\wedge\varepsilon^2=
\]
\[
=-{\bf d}({\boldsymbol{E}}\wedge\boldsymbol{B})\otimes\varepsilon^1\wedge\varepsilon^2=
-{\bf d}(**({\boldsymbol{E}}\wedge\boldsymbol{B}))\otimes\varepsilon^1\wedge\varepsilon^2=
*\delta({\boldsymbol{E}}\times\boldsymbol{B})\otimes\varepsilon^1\wedge\varepsilon^2=
\]
\[
=-*{\rm div}({\boldsymbol{E}}\times\boldsymbol{B})\otimes\varepsilon^1\wedge\varepsilon^2=
-{\rm div}({\boldsymbol{E}}\times\boldsymbol{B})dx\wedge dy\wedge dz\otimes
\varepsilon^1\wedge\varepsilon^2.
\]
Following the same rules we obtain
\[
\wedge\left(\omega,*\frac1c\frac{\partial }{\partial t}
{\cal I}_*\omega\right)=
\frac1c\frac{\partial }{\partial t} \frac{{\boldsymbol{E}}^2+
\boldsymbol{B}^2}{2}dx\wedge dy\wedge
dz\otimes\varepsilon^1\wedge\varepsilon^2, \] and \[
\frac{4\pi}{c}\wedge(\omega,*{\cal I}_*{\cal J})=
\frac{4\pi}{c}\left({\boldsymbol{E}}.\boldsymbol{j}_e-
\boldsymbol{B}.\boldsymbol{j}_m\right) dx\wedge dy\wedge dz
\otimes\varepsilon^1\wedge\varepsilon^2.
\]
Hence, the generalized Poynting energy-momentum
balance relation is given by
\begin{equation}
\wedge\left(\omega, {\bf d}\omega-*\frac1c\frac{\partial }{\partial t}
{\cal I}_*\omega\right)=\frac{4\pi}{c}\wedge\left(\omega,{\cal I}_*{\cal
J}\right).                                  %32%
\end{equation}
As for the generalized Lorentz force $\vec{\cal F}$, staying on the
right-hand side of eq.(13), it is presented by
\begin{equation}
\vec{\cal F}\otimes\varepsilon^1\wedge\varepsilon^2=
\left[\frac1c\left(\boldsymbol{j}_e\times\boldsymbol{B}-\boldsymbol{j}_m\times{\boldsymbol{E}}\right)+
\rho_e{\boldsymbol{E}}+\rho_m\boldsymbol{B}\right]\otimes\varepsilon^1\wedge\varepsilon^2=
\frac1c*\wedge\left(\omega,{\cal J}\right)+              %33%
*\wedge\left(\omega,{\cal I}_*{\cal Q}\right).
\end{equation}
Since the orthonormal 2-form $\varepsilon^1\wedge\varepsilon^2$ is invariant
with respect to rotations (and even with respect to unimodular
transformations in ${\cal R}^2$) we have the duality invariance of the
above energy-momentum quantities and relations.

Hence, in our approach we have achieved a full covariance of the equations,
given in the form (31).  Indeed, the covariance with respect to arbitrary
transformations in ${\cal R}^3$ is obvious, so we show now the covariance of
(31) with respect to nonsingular linear transformations $\varphi:{\cal
R}^2\rightarrow {\cal R}^2$. Let $\omega$ satisfies (31), so we have to show
that $\varphi_*(\omega)= {\boldsymbol{E}}\otimes\varphi(\varepsilon^1)
+\boldsymbol{B}\otimes\varphi(\varepsilon^2)$ also satisfies (31). We apply
$\varphi$ from the left on (31) and obtain
\[
\varphi\left(*{\bf d}\omega-
\frac1c\frac{\partial }{\partial t}{\cal I}_*\omega\right)=
\left(*{\bf d}\varphi_*\omega-
\frac1c\frac{\partial }{\partial t}\varphi_*{\cal I}_*\omega\right)=
\left(*{\bf d}\varphi_*\omega-
\frac1c\frac{\partial }{\partial t}\varphi_*{\cal
I}_*\varphi^{-1}_*\varphi_*\omega\right)=
\]
\[
=\left(*{\bf d}\varphi_*\omega-
\frac1c\frac{\partial }{\partial t}(\varphi{\cal I}
\varphi^{-1})_*\varphi_*\omega\right)=
\left(*{\bf d}-
\frac1c\frac{\partial }{\partial t}{\cal I}'_*\right)
\varphi_*\omega,
\]
where ${\cal I}'=\varphi{\cal I}\varphi^{-1}$. For the right-hand side of
(31) we obtain
$$
(\varphi{\cal I})_*{\cal J}=(\varphi{\cal I}\varphi^{-1})_*
\varphi_*{\cal J}={\cal I}'_*\varphi_*{\cal J},
$$
so, our assertion is proved.

Note the following simple forms of the energy density
\[
\frac{1}{8\pi}*\wedge\left(\omega,*{\cal I}_*\omega\right)=
\frac{{\boldsymbol{E}}^2+\boldsymbol{B}^2}{8\pi}\varepsilon^1\wedge\varepsilon^2,
\]
and of the Poynting vector,
\[
\frac{c}{8\pi}*\wedge(\omega,\omega)=
\frac{c}{4\pi}{\boldsymbol{E}}\times\boldsymbol{B}\otimes\varepsilon^1\wedge\varepsilon^2,
\]
the D-invariance is obvious. As for the general ${\cal R}^2$ covariance of
the second equation of (31) it is obvious.

Resuming, we may say that pursuing the correspondence: {\it one physical
object - one mathematical model-object}, we came to the idea to introduce the
${\cal R}^2$-valued 1-form $\omega$ as the mathematical model-field.  This,
in turn, set the problem for general ${\cal R}^2$ covariance of the equations
and this problem was solved through introducing explicitly the canonical
complex structure ${\cal I}$ in the dynamical equations (31) of the theory.
The duality transformation appears now as an invariance property of the
energy-momentum quantities and relations.

\vskip 0.5cm

{\bf 3.2 \ Relativistic formulation}. As it was mentioned in the previous
section in the relativistic formulation of Maxwell equations we have got
already a general ${\cal R}^2$-covariance, but the two components, subject to
the general ${\cal R}^2$-linear transformation, are {\it not} independent, in
fact they are $({\bf F},*{\bf F})$. We look now at the situation from another
point of view.

First we note that we shall follow the main idea of the nonrelativistic
formulation, namely, to consider as a mathematical-model
field some ${\cal R}^2$-valued differential form, being closely connected
to the canonical complex structure ${\cal I}$ in ${\cal R}^2$. But in
contrast to the nonrelativistic case here we consider an ${\cal R}^2$-valued
2-form on the Minkowski space-time $M=({\cal R}^4,\eta)$.  In general such a
2-form $\Omega$ looks as $\Omega={\bf F}_1\otimes\varepsilon^1+{\bf
F}_2\otimes \varepsilon^2={\bf F}_a\otimes \varepsilon^a $. Let now we are
given two linear maps:
$$
\Phi:\Lambda^2(M)\rightarrow\Lambda^2(M),
$$
$$
\varphi:{\cal R}^2\rightarrow{\cal R}^2.
$$
These maps induce a map
$(\Phi,\varphi):\Lambda^2(M,{\cal R}^2)\rightarrow\Lambda^2(M,{\cal R}^2)$
by the rule:
\[
(\Phi,\varphi)(\Omega)=(\Phi,\varphi)({\bf F}_a\otimes\varepsilon^a)=
\Phi({\bf F}_a)\otimes\varphi(\varepsilon^a).
\]
It is natural to ask now is it possible the joint action of these two maps to
keep $\Omega$ unchanged, i.e. to have
$$
(\Phi,\varphi)(\Omega)=\Omega.
$$
In such a case the form $\Omega$ is called
$(\Phi,\varphi)$-{\it equivariant}.  If $\varphi$ is a linear isomorphism and
we identify $\Phi$ with $(\Phi,id)$ and $\varphi$ with $(id,\varphi)$, we can
equivalently write
\[
\Phi(\Omega)=\varphi^{-1}(\Omega).
\]
If we specialize now: $\varphi={\cal I}$ we readily find that
the $(\Phi,{\cal I})$-equivariant forms $\Omega$ must satisfy
\[
(\Phi,{\cal I})(\Omega)=
-\Phi({\bf F}_2)\otimes\varepsilon^1+\Phi({\bf F}_1)\otimes\varepsilon^2=
{\bf F}_1\otimes\varepsilon^1+{\bf F}_2\otimes\varepsilon^2,
\]
Hence, we must have $\Phi({\bf F}_1)={\bf F}_2$ and $\Phi({\bf
F}_2)=-{\bf F}_1$, i.e. $\Phi^2=-id$.
In other words, the property ${\cal I}^2=-id$ is carried over to $\Phi$:
$\Phi^2=-id$. Since the Hodge $*$, restricted to 2-forms in Minkowski space,
satisfies this condition, and according to expressions (19)-(20) in standard
coordinates its action coincides with the special duality transformation, it
is a natural candidate for $\Phi$. Hence, working with $(*,{\cal
I})$-equivariant 2-forms on Minkowski space, we can replace the action of
${\cal I}$ with the action of the Hodge $*$-operator. And that's why in
relativistic electrodynamics we have general ${\cal R}^2$ covariance if we
work with forms $\Omega$ of the kind
$\Omega={\bf F}\otimes\varepsilon^1+*{\bf F}\otimes\varepsilon^2$.
In the nonrelativistic formulation this is not possible to be done since we
work there with 1-forms on ${\cal R}^3$ and no map $\Phi:\Lambda^1({\cal
R}^3)\rightarrow\Lambda^1({\cal R}^3)$ with the property $\Phi^2=-id$ exists,
and we have to introduce the complex structure through ${\cal R}^2$ only.

Having in view these considerations our basic assumption for the
algebraic nature of the mathematical-model object must read:
\vskip 0.5cm
{\it The electromagnetic field is (relativistically) represented by a}
$(*,{\cal I})$-{\it equivariant 2-form} $\Omega$ {\it on the Minkowski
space-time}:
\begin{equation}
\Omega={\bf F}\otimes\varepsilon^1+*{\bf F}\otimes\varepsilon^2.      %34%
\end{equation}
We note two algebraic properties of the forms of the kind (34).
First, all such $p$-forms, where the basis is fixed, form a linear space.
Further, if $\Omega$ is of the form (34) with respect to some basis
$(\varepsilon^1,\varepsilon^2)$, i.e. if $(*,{\cal I})(\Omega)=\Omega$,
this does not mean that it will have the same form with respect to any other
basis $(e_1,e_2)$ of ${\cal R}^2$.
But if these two bases are transformed into
each other through a symmetry $S$ of ${\cal I}$, i.e. If
$S{\cal I}S^{-1}={\cal I}$, we shall prove that $\Omega$ will have the same
form (38) with respect to $(e_1,e_2)$.  We shall use the notation:
$\Omega_\varepsilon\equiv {\bf F}\otimes
\varepsilon^1+*{\bf F}\otimes\varepsilon^2$.
Now let
$\varepsilon^1=S(e^1)\rightarrow e^1=S^{-1}(\varepsilon^1),\
\varepsilon^2=S(e^2)\rightarrow e^2=S^{-1}(\varepsilon^2);\
S{\cal I}S^{-1}={\cal I}$.
The last relation is equivalent to any of the following two relations:
$S^{-1}{\cal I}S={\cal I},\ S^{-1}{\cal I}={\cal I}S^{-1}$.
We have also  ${\cal I}(\varepsilon^1)=\varepsilon^2$ and
${\cal I}(\varepsilon^2)=-\varepsilon^1$.
So, we have to prove the relation:
$(*,{\cal I})(\Omega_e)=\Omega_e$, where
$\Omega_e=(id,S^{-1})(\Omega_\varepsilon)$. First we prove
${\cal I}(e^1)=e^2$ and ${\cal I}(e^2)=-e^1$:
\[
S{\cal I}S^{-1}(\varepsilon^1)={\cal I}(\varepsilon^1)=\varepsilon^2=
S{\cal I}(e^1)\rightarrow S^{-1}(\varepsilon^2)=e^2={\cal I}(e^1)
\rightarrow {\cal I}(e^2)=-e^1.
\]
Further,
\[
(*,{\cal I})\Omega_e=(*,{\cal I}S^{-1})\Omega_\varepsilon
=(*,S^{-1}{\cal I})\Omega_\varepsilon=
(*,S^{-1})({\bf F}\otimes \varepsilon^2-*{\bf F}\otimes\varepsilon^1)=
\]
\[
=*{\bf F}\otimes S^{-1}(\varepsilon^2)-
**{\bf F}\otimes S^{-1}(\varepsilon^1)=
*{\bf F}\otimes e^2+{\bf F}\otimes e^1=\Omega_e.
\]
This means that the set of all $(*,{\cal I})$-equivariant forms is partitioned
to nonoverlaping subclasses with respect to the action of the automorphisms
of the complex structure ${\cal I}$.

The relativistic pure field Maxwell equations (18), expressed through the
\linebreak $(*,{\cal I})$-equivariant 2-form $\Omega$
have, obviously, general ${\cal R}^2$ covariance and are equivalent to
\begin{equation}
{\bf d}\Omega=0.                               %35%
\end{equation}
In presence of electric and magnetic charges, making use of the definitions
in the previous section, we introduce the generalized relativistic current
${\cal J}_r$ as follows
$$
{\cal J}_r=j_e\otimes\varepsilon^1+J_m\otimes\varepsilon^2.
$$
Hence, equations (25)-(26) acquire the form
\begin{equation}
{\bf d}\Omega=-4\pi{\cal J}_r.                     %36%
\end{equation}
The generalized Lorentz force is given by
\begin{equation}
-*\wedge({\cal J}_r,\Omega)=
\Bigl(-{\bf F}_{\mu\sigma}j_e^\sigma -                    %37%
(*{\bf F})_{\mu\sigma}J_m^\sigma \Bigr)dx^\mu
\otimes\varepsilon^1\wedge\varepsilon^2.
\end{equation}
The divergence of the energy-momentum tensor ${\bf Q}_\mu^\nu$ is given by
\begin{equation}
*\wedge\Bigl(\delta\Omega, \Omega\Bigr)=
-\frac{1}{4\pi}\Bigl[{\bf F}_{\mu\nu}\nabla_\sigma {\bf F}^{\sigma\nu}+
(*{\bf F})_{\mu\nu}\nabla_\sigma (*{\bf F})^{\sigma\nu}\Bigr]dx^\mu\otimes
\varepsilon^1\wedge\varepsilon^2.
\end{equation}
The energy-momentum tensor ${\bf Q}$, considered as a symmetric 2-form
${\bf Q}_{\mu\nu}={\bf Q}_{\nu\mu}$, is given by
\begin{equation}
\left({\bf Q}\otimes\varepsilon^1\wedge\varepsilon^2\right)(X,Y)=        %38%
\frac{1}{8\pi}*\wedge\Bigl(i_X\Omega,*i_Y{\cal I}_*\Omega\Bigr),
\end{equation}
where $X$ and $Y$ are 2 arbitrary vector fields, and $i_X$ is the inner
product by the vector field $X$. Indeed,
\[
i_X\Omega=X^\mu {\bf F}_{\mu\nu}dx^\nu\otimes\varepsilon^1+
X^\mu (*{\bf F})_{\mu\nu}dx^\nu\otimes\varepsilon^2,\
\]
\[
*i_Y{\cal I}_*\Omega=
*\Bigl[Y^\mu {\bf F}_{\mu\nu}dx^\nu\Bigr]\otimes\varepsilon^2-
*\Bigl[Y^\mu (*{\bf F})_{\mu\nu}dx^\nu\Bigr]\otimes\varepsilon^1.
\]
\[
\wedge\Bigl(i_X\Omega,*i_Y{\cal I}_*\Omega\Bigr)=
-X^\mu Y^\nu\Bigl[{\bf F}_{\mu\sigma}{\bf F}_\nu^\sigma +
(*{\bf F})_{\mu\sigma}(*{\bf F})_\nu^\sigma\Bigr]dx\wedge dy\wedge dz\wedge
d\xi \otimes\varepsilon^1\wedge\varepsilon^2.
\]
So, we obtain
\[
\frac{1}{8\pi}*\wedge(i_X\Omega,*i_Y{\cal I}_*\Omega)=
-\frac{1}{8\pi}X^\mu Y^\nu\Bigl[{\bf F}_{\mu\sigma}F_\nu^\sigma +
(*{\bf F})_{\mu\sigma}(*{\bf F})_\nu^\sigma\Bigr]
\varepsilon^1\wedge\varepsilon^2.
\]
The presence of the 2-form $\varepsilon^1\wedge\varepsilon^2$ introduces
invariance with respect to unimodular transformations in ${\cal R}^2$.

Another definition of the energy-momentum tensor, making no use of the
complex structure ${\cal I}$, is through the
canonical inner product $g$ in ${\cal R}^2$ as a bilinear map instead of
$\wedge$.  Indeed, it is easy to see that the right-hand side of the relation
\[
{\bf Q}_{\mu\nu}X^\mu Y^\nu=\frac12 *g\Bigl(i(X)\Omega,*i(Y)\Omega\Bigr)
\]
is equal to
\[
-\frac12 X^\mu Y^\nu\biggl[{\bf F}_{\mu\sigma}{\bf F}_\nu^\sigma +
(*{\bf F})_{\mu\sigma}(*{\bf F})_\nu^\sigma\biggr].
\]

Resuming, we note the main differences with respect to the nonrelativistic
formulation. First, the mathematical model-object is a 2-form $\Omega$ on
Minkowski space with values in ${\cal R}^2$, second, $\Omega$ is
$(*,{\cal I})$-equivariant. As for the usual duality transformations, they
appear as particular ${\cal R}^2$-invariance properties of the conserved
quantities and of the corresponding conservation relations.

\vskip 0.5cm
The general conclusion of this section is that the ${\cal R}^2$ valued
nonrelativistic 1-form $\omega$ and the relativistic 2-form $\Omega$ seem to
be natural and adequate mathematical model-objects of electromagnetic fields,
while the duality transformations characterize the invariance properties of
the conversed quantities and the corresponding conservation relations.

\section{Conclusion}
Here we are going to mention those points of the paper which from our
point of view seem most important.

Classical electrodynamics works mainly with two concepts: {\it charge} and
{\it field}. The charge carriers (called also field sources) are considered
as point-like (or structureless) objects.  The field is considered as
generated by static or moving charges, and it is not defined at the points of
its own source, hence, the point charges acquire topological sense. Passing
to continuous charge distributions we write down currents on the right hand
sides of Maxwell equations and forget about the topological nature of
charges. Moreover, this identification of characteristics of the field
represented by the left hand sides of the equations with characteristics of
mass objects carrying electric (and possiply magnetic) charges, seems not
well enough motivated from theoretical point of view. It would be more
natural to write down equations which identify quantities of the {\it same}
nuture, e.g. some energy-momentum balance relation between the field and the
particles (recall our remark at the end of Sec.2).

The duality properties of the solutions reveal the internal structure of the
field as having {\it two vector components}, which are

-differentially interrelated (through the equations), but

-algebraically distinguished.

Moreover, the adequate understanding of the duality properties requires
explicitly introduced complex structure in the equations.

This resulted in making use of ${\cal R}^2$-valued differential forms,
$\omega$ and $\Omega$, as mathematical model objects, and corresponding
complex structures ${\cal I}$ and the relativistic Hodge $*$ restricted to
2-forms.  The equations admit a full ${\cal R}^2$ covariance, while the
duality properties appear as invariance properties of the conservative
quantities and conservation relations. In fact, the action of the
D-transformations in the linear space of vacuum solutions  separates
classes of solutions with the same energy-momentum.

Finally, we may expect that recognizing the structure of the field as
a {\it double vector-component} one through its {\it duality} properties may
open new ways of considerations and may generate new ideas and developments
towards an appropriate nonlinearization of classical electrodynamics.

\newpage
\vskip 1cm
{\bf REFERENCES}
\vskip 0.5cm
1. {\bf Olive, D.},hep-th/9508089; {\bf Harvey, J.},hep-th/9603086;
   {\bf Gomez, C.},hep-th/9510023; \newline{\bf Verlinde, E.},hep-th/9506011

2. {\bf Heaviside, O.}, "Electromagnetic Theory" (London, 1893); Electrical
Papers, London, {\bf 1} (1892); Phyl.Trans.Roy.Soc., 183A, 423 (1893)

3. {\bf Larmor, J.}, Collected Papers. London, 1928

4. {\bf Rainich, G.}, Trans.Amer.Math.Soc., {\bf 27}, 106 (1925)

5. {\bf Misner, C.}, {\bf Wheeler, J.} Ann. of Phys., {\bf 2}, 525 (1957)

6. "The Dirac Monopole", (Mir, Moscow, 1970)

7. {\bf Strajev, V.}, {\bf Tomilchik, L.}, "Electrodynamics with Magnetic
Charge", (Nauka i Tehnika, Minsk, 1975)

8. {\bf Gibbons, G.}, {\bf Rasheed, D.}, hep-th/9506035

9. {\bf Chruscinski, D.}, hep-th/9906227

10. {\bf Chan Hong-Mo}, hep-th/9503106; hep-th/9512173

\end{document}